\documentclass[aps,prl,twocolumn,showpacs]{revtex4}

\usepackage{graphicx}%
\usepackage{dcolumn}
\usepackage{amsmath}

\begin{document}
\title{Bose-Einstein condensation at constant temperature}
\author{M.~Erhard}
\author{H.~Schmaljohann}%
\author{J.~Kronj\"ager}%
\author{K.~Bongs}%
\author{K.~Sengstock}%
\affiliation{%
Institut f\"ur Laser-Physik,
Universit\"at Hamburg,
Luruper Chaussee 149,
22761 Hamburg,
Germany}
\date{\today}
\begin{abstract}
We present a novel experimental approach to Bose-Einstein condensation by increasing the particle number of the system at almost constant temperature. In particular the emergence of a new condensate is observed in multi-component $F\!=\!1$ spinor condensates of $^{87}$Rb. Furthermore we develop a simple rate-equation model for multi-component BEC thermodynamics at finite temperature which well reproduces the measured effects.
\end{abstract}
\pacs{03.75.Nt, 03.75.Hh, 03.75.Mn}
\maketitle
\begin{figure}[htbp]
  \centering
  \includegraphics[width=86mm]{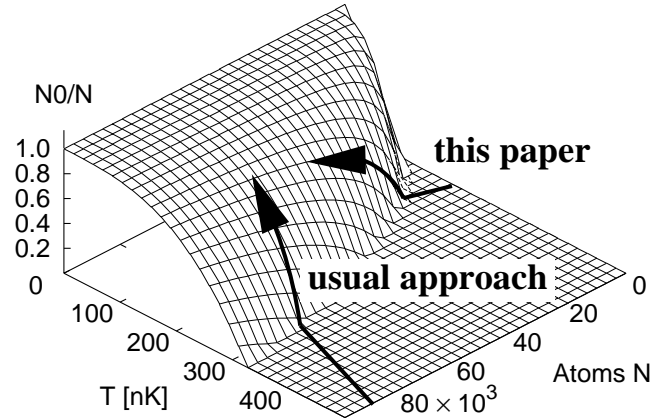}
  \caption{Phase diagram of Bose-Einstein condensation for a typical $^{87}$Rb experiment. The condensate fraction (if $>\!0$) is plotted as $N_0/N=1-g_3(1)(k_B/(\hbar\bar{\omega}))^3T^3/N$. The usual realization of BEC is done by decreasing $T$ at (almost) constant particle number $N$. In this paper condensation by increasing particle number starting with $N=0$ at (nearly) constant temperature is discussed.}
  \label{fig:1}
\end{figure}

The experimental realization of Bose-Einstein condensates (BEC) in dilute atomic gases \cite{Bradley1995a,Davis1995b,Anderson1995a} and the breathtaking emergence of fascinating physics of cold quantum gases in an increasing number of experiments have had formative influence on the common model usually used for the description of Bose-Einstein condensation (see e.g.~\cite{Pitaevskii2003a} and references therein). This model is based on a system of constant particle number whose temperature $T$ is reduced. The popularity of this approach arises from the fact that all experiments so far make use of evaporative cooling techniques which reduce the temperature of the sample (at the expense of particle losses). This path to quantum degeneracy is illustrated in the phase diagram of Fig.~\ref{fig:1}. Starting with a certain particle number $N$, the temperature $T$ of the system is reduced below the critical temperature $T_c(N)$ which leads to an accumulation of particles in the condensate fraction $N_0/N$. Detailed experimental studies \cite{Ensher1996a,Mewes1996a,Gerbier2004a} have compared this quantity with theoretical descriptions.
  
In this paper we present a completely different and new experimental realization of Bose-Einstein condensation by increasing the particle number of a system at almost constant temperature. The corresponding path is also marked in Fig.~\ref{fig:1} and leads to BEC almost orthogonally to the common route discussed above. We start with $N=0$ and add more and more particles at nearly constant temperature $T$ until the critical particle number $N_c(T)$ is reached, i.e.~the population of the thermal cloud saturates and all further added particles fill up the condensate fraction. Worth mentioning this approach corresponds to the original idea used by Einstein \cite{Einstein1925a} and theoretical descriptions over decades to discuss BEC. Furthermore first attempts to achieve quantum degeneracy in spin-polarized hydrogen \cite{Silvera1980a,Cline1980a} were based on increasing density by adding particles and by compression at liquid helium temperatures.

The new thermodynamical approach to BEC discussed in this letter is realized in multi-component BEC which provide multiple internal quantum states of the involved atoms. We want to emphasize that these systems open up a rich variety of new thermodynamical aspects as the involved finite temperature dynamics is extended to more components which are additionally coupled and influence each other. The thermodynamical description has to take into account all interactions between multiple condensate components and just as many thermal clouds (we use this term instead of 'normal components').
In this context recent experiments  have observed 'decoherence-driven cooling' \cite{Lewandowski2003a} and melting of new condensate components \cite{Schmaljohann2004a}. 

\begin{figure}[htbp]
  \centering
  \includegraphics[width=60mm]{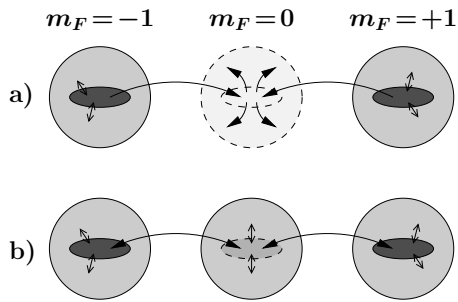}
  \caption{Scheme of the dynamics. (a) Spin dynamics transfers population to the $m_F=0$ state which thermalizes almost immediately. (b) When all thermal clouds are equally populated and thus the critical particle number in $m_F\!=\!0$ is reached a condensate arises and 'free' spin dynamics can take place.}
  \label{fig:2}
\end{figure}

The system considered here is based on a $F\!=\!1$ spinor condensate of $^{87}$Rb with three internal states $m_F=-1,0,+1$. The main idea is to increase the particle number in the initially unpopulated $m_F\!=\!0$ spin component via spin dynamics transfer out of the other components. For this we first prepare a partially condensed mixture of the $|\!-1\rangle$ and $|\!+1\rangle$ states. 
The resulting dynamics can be divided into two main successive steps which are illustrated in Fig.~\ref{fig:2} as a) and b).

The first process is that spin dynamics populates the $m_F\!=\!0$ state by converting $m_F=\pm 1$ condensate atoms into $m_F\!=\!0$ atoms according to $|\!+1\rangle + |\!-1\rangle \leftrightarrow |0\rangle + |0\rangle $ \cite{Ohmi1998a,Ho1998a,Stenger1999a,Schmaljohann2004a}.
 Due to its density dependence spin dynamics is practically restricted to the condensed fractions, resulting in the production of $m_F\!=\!0$ 'condensate' atoms, which however immediately thermalize into the $m_F\!=\!0$ thermal cloud due to collisions with all thermal clouds (Fig.~\ref{fig:2}a). We want to emphasize at this point that thermalization is the fastest timescale ($\approx$\,50\,ms) of our system and therefore spin dynamics ($\approx$\,1\,s) is only a means to produce the new component. 
The redistribution of constant total energy among more thermal atoms during this process leads as a side-effect to a decrease of temperature $T$. This is similar to 'decoherence-driven cooling' of the JILA experiment \cite{Lewandowski2003a} which in contrast to our system did not involve conversion between different condensate components.

As soon as the critical particle number in the $m_F\!=\!0$ thermal cloud is reached the phase transition in the $m_F\!=\!0$ component takes place and a condensate fraction emerges (Fig.~\ref{fig:2}b). From this moment on the thermal clouds are and remain equally populated and provide a constant temperature reservoir of the system. Therefore 'free' spin dynamics may take place between the spin components of the condensate fractions, i.e. at constant total number of condensed atoms but still in touch with the reservoir of finite temperature. Thus spin dynamics mainly determines the final $m_F$ condensate fractions, which are not as a rule equally populated in contrast to the thermal clouds \cite{Erhard2004g}. 

%
%
%
The experimental setup (for details see \cite{Schmaljohann2004a,Erhard2003a}) produces BECs in an optical dipole trap which provides a spin-independent trapping potential. The trapping frequencies are $2\pi\times 890$\,Hz vertically, $2\pi\times 160$\,Hz horizontally and $2\pi\times 20$\,Hz along the beam direction. 
Spin dynamics is suppressed during preparation of the initial spin state due to the high magnetic offset field of 25\,G which is subsequently lowered to a value of $340\pm20$\,mG to allow for spin dynamics. After a variable hold time of 0..30\,s the dipole trap is switched off and the released atoms are spatially separated by a Stern-Gerlach gradient. Finally an absorption image is taken in order to determine BEC and thermal atom numbers by a simultaneous fit of three parabolas and three Gaussians for the three $m_F$ components.

\begin{figure}[htbp]
  \centering
  \includegraphics[width=86mm]{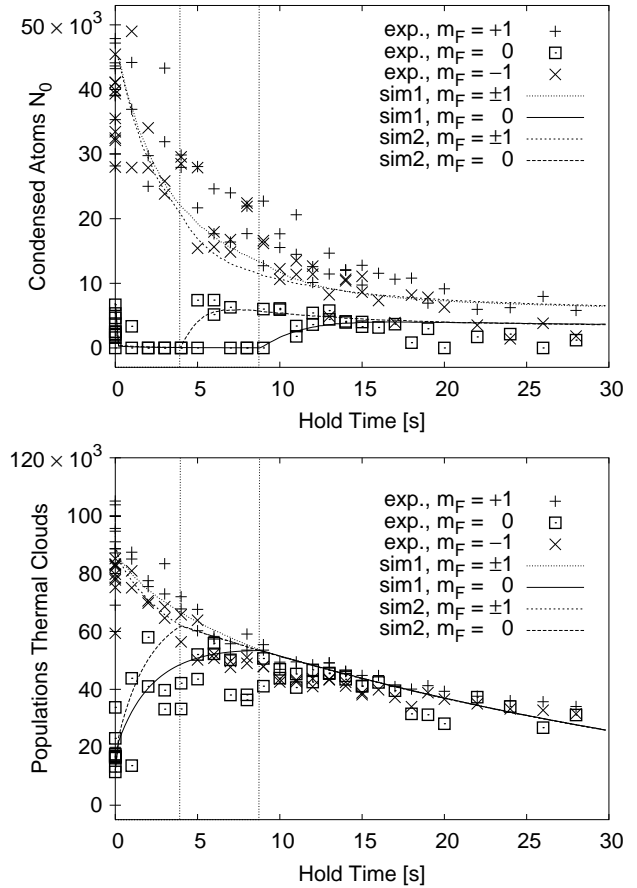}
  \caption{Measured condensate and thermal atom numbers for the different spin states (marked as 'exp.') as function of different hold times. The lines represent solutions of the rate-equation model for two different sets of spin dynamics parameters denoted as 'sim1' and 'sim2' (see text for numbers). The moments when the critical particle number for $m_F\!=\!0$ is reached in the simulations are marked by vertical lines.}
  \label{fig:3}
\end{figure}
Fig.~\ref{fig:3} shows the experimentally obtained BEC and thermal atom numbers versus the hold time compared to simulations of the rate-equation model which will be presented later. We start with an initial mixture of $m_F=\!\pm\!1$ both in BEC and thermal fractions. The preparation process leads to a remaining population of $<10$\,\% in the $m_F\!=\!0$ state.

The experimental data demonstrate all of the previously introduced dynamics only modified by loss processes. First a $m_F\!=\!0$ thermal cloud arises and grows until the critical particle number is reached after 5..10\,s. Note, this is the moment of equal populations of all thermal clouds. Subsequently a $m_F\!=\!0$ condensate fraction emerges. The data between 5..10\,s suggest that the exact moment of phase transition varies from shot to shot. Indeed this moment crucially depends on spin dynamics as will be discussed later. 
Finally spin dynamics leads to a steady-state which decreases due to loss processes with an experimentally observed relative condensate distribution of 40..45\,\% $m_F=\!\pm 1$ and 10..20\,\% $m_F\!\!=\!0$ \cite{Erhard2004f}.

%
In the following we develop a simple rate-equation model which reproduces the main experimental observations. We do not intend to give a detailed and thorough simulation of finite temperature BEC which would be quite involved and is subject of current theoretical activities \cite{Gardiner2000a,Morgan2003a,Zaremba2002a,Goral2002a}. Rather a basic model from a experimentalist's point of view is presented to stimulate a vivid discussion of finite temperature effects in multi-component BECs and introduce a number of single processes which yield the observed behavior. The model is based on a set of 7 variables ${N_0^-,N_0^0,N_0^+,N_t^-,N_t^0,N_t^+,T }$ where $N_0^X$ with $X=-,0,+$ denote the atom numbers of $m_F=-1,0,+1$ in the condensate fraction and $N_T^X$ the respective atom numbers in the thermal cloud. $T$ is the system temperature and assumed to be equal for all components.
The equations of motion read 
\begin{eqnarray*}
  \dot{N}_0^X &=& \dot{N}_{0,th}^X + \dot{N}_{0,sp}^X + \dot{N}_{0,1b}^X + \dot{N}_{0,3b}^X \quad, \\
  \dot{N}_t^X &=& \dot{N}_{t,th}^X + \dot{N}_{t,1b}^X + \dot{N}_{t,ev}^X \quad, \\
  \dot{T} &=& \dot{T}_{th} + \dot{T}_{ev}\quad,
\end{eqnarray*}
and include the processes thermalization ($\dot{N}_{\star,th}^X$, $\dot{T}_{th}$), spin dynamics ($\dot{N}_{0,sp}^X$), one-body losses ($\dot{N}_{\star,1b}^X$), three-body losses ($\dot{N}_{0,3b}^X$) and evaporation ($\dot{N}_{t,ev}^X$, $\dot{T}_{ev}$). These single effects as well as an aditionally introduced phase-space redistribution will be discussed in the following.

The thermalization rate $\gamma_{th}$ quantifies the collisional transfer of condensed atoms into the thermal component
\begin{eqnarray*}
  \dot{N}_{0,th}^X &=& -\tilde{\gamma}_{th} N_0^X N_t \quad,\\
  \dot{N}_{t,th}^X &=& +\tilde{\gamma}_{th} N_0^X N_t \quad,
\end{eqnarray*}
where $N_t=N_t^- + N_t^0 + N_t^+$ and $\tilde{\gamma}_{th}$ is obtained via the relation $\tilde{\gamma}_{th} N_t = \gamma_{th} \hat{n}_t$ which takes into account the peak density of the thermal cloud $\hat{n}_t$ to convert the density dependent rate $\gamma_{th}$ into $\tilde{\gamma}_{th}=\gamma_{th} \bar{\omega}^3 (m/(2\pi k_B T))^{(3/2)}$. The temperature- and spin-dependence of $\gamma_{th}$ is neglected. The system temperature $T$ decreases as the conserved total energy is redistributed among more thermal atoms and given as
\begin{equation*}
  \dot{T}_{th} = - T \tilde{\gamma}_{th} N_0 \quad,
\end{equation*}
with $N_0=N_0^- + N_0^0 + N_0^+$.
The used value $\gamma_{th}=10^{-18}$\,m$^3$/s leads to a thermalization rate $\tilde{\gamma}_{th} N_0$ of $\approx 13$ 1/s for $N_0=45000$ which corresponds to our experiment.

Spin dynamics is implemented by a simple coupling of the condensate atoms due to the relation $|\!-1\rangle + |\!+1\rangle \leftrightarrow |0\rangle + |0\rangle$ with two reaction rates $\tilde{\gamma}_{sp1}$ and $\tilde{\gamma}_{sp2}$ for forward and backward reaction \cite{Erhard2004a}:
\begin{eqnarray*}
  \dot{N}_{0,sp}^{\pm} &=& \tilde{\gamma}_{sp1} N_0^0 N_0^0  -
                           \tilde{\gamma}_{sp2} N_0^- N_0^+ \quad ,\\
  \dot{N}_{0,sp}^0     &=& -2\tilde{\gamma}_{sp1} N_0^0 N_0^0  +
                           2\tilde{\gamma}_{sp2} N_0^- N_0^+ 
                        \quad\mbox{.}
\end{eqnarray*}

One-body loss occurs with rate $\gamma_1$ independently of the spin state and equally in the BEC and thermal cloud. The value used is $\gamma_1=0.011$ 1/s and corresponds to the measured 1/e-lifetime of 90\,s limited by background gas collisions
\begin{eqnarray*}
  \dot{N}_{0,1b}^X &=& -\gamma_1 N_0^X \quad ,\\
  \dot{N}_{t,1b}^X &=& -\gamma_1 N_t^X  \quad .
\end{eqnarray*}
   
For three-body loss \cite{Soding1999a} we assume a spin-independent process, ignore possible changes in statistical factors due to multiple components and obtain
\begin{equation*}
  \frac{\dot{N}_{0,3b}^X}{N_0^X} = -L c_3 (N_0)^{4/5} \quad,
\end{equation*}
with $c_3=7/6 c_2^2$ and $c_2=15^{2/5}(14\pi)^{-1}(m\bar{\omega}/\hbar\sqrt{a})^{6/5}$. The loss rate used is $L=5.8\times 10^{-42}$\,m$^6$/s \cite{Burt1997a}.

The evaporation process due to finite trap depth $k_B T_e$ is implemented by a particle loss of the thermal cloud connected with a decrease of the system temperature.
The temperature dependence of the evaporation rate $\gamma_e$ is neglected and the loss reads
\begin{equation*}
  \dot{N}_{t,ev}^X = -\gamma_e N_t^X\quad \mbox{.}
\end{equation*}
Energy conservation leads to a change of temperature
\begin{equation*}
  \dot{T}_{ev} = \gamma_e (T-T_e) \quad\mbox{.}
\end{equation*}

We use the Euler method to propagate the equations in discrete time-steps of duration $\Delta t$ (e.g.~$N_0^X(t+\Delta t) = N_0^X(t) + \dot{N}_0^X\Delta t$).
After each simulation step a phase-space redistribution is carried out. This is important to introduce quantum statistics into the equations and can be regarded as spontaneous condensation if the critical density is reached. The critical particle number is calculated as $N_c = g_3(1)(k_B T/(\hbar\bar{\omega}))^3$ and the following condition is checked for $X=-,0,+$
\begin{eqnarray*}
  \mbox{If\,\,} (N_t^X > N_c) &:& N_0^X(t+\Delta t) = N_0^X(t) + (N_t^X(t) - N_c)\\
                       && N_t^X(t+\Delta t) = N_c \quad\mbox{.}
\end{eqnarray*}
This re-condensation step is related to a temperature change obtained by total energy conservation as
\begin{equation*}
  T(t+\Delta t) = T(t) \left( 1+\frac{N_0(t+\Delta t)-N_0(t)}{N_t(t+\Delta t)} \right)
       \quad\mbox{.}
\end{equation*}
The thermalization step and the phase-space redistribution cancel out in the case of thermal equilibrium resulting in steady condensate fractions and constant temperature. Nevertheless these steps are crucial to describe the occurrence of the new thermal components and new condensate fractions. 
As thermalization is the fastest timescale of the considered system a step-like description seems to be reasonable.

%
%
Our rate-equation model reproduces all experimentally observed thermal features even with a reasonable quantitative accuracy as shown in Fig.~\ref{fig:3}.
The initial condensate populations were chosen as $N_0^- (0) = N_0^+ (0) = 45000$, $N_0^0 (0) = 7000$ and the thermal atom numbers as $N_t^- (0) = N_t^+ (0) = 90000$ and $N_t^0 (0) = 12000$ and $T(0)=288$\,nK. 
Evaporation parameters are $\gamma_e = 0.015$ 1/s and $T_e = 500$\,nK.
Simulations for two sets of spin rates $\tilde{\gamma}_{sp1}=1.6\times 10^{-5}$\,1/s, $\tilde{\gamma}_{sp2}=0.4\times 10^{-5}$\,1/s (sim1) and $\tilde{\gamma}_{sp1}=2.4\times 10^{-5}$\,1/s, $\tilde{\gamma}_{sp2}=0.6\times 10^{-5}$\,1/s (sim2) have been carried out. Although these two sets of parameters differ by only 33\,\% the resulting moment of condensation varies by more than a factor of two (4 and 9\,s respectively). Indeed we have to assume that there is a shot-to-shot variation of spin dynamics in our experiment as initial phases are not controlled. It has been theoretically shown \cite{Law1998c} that spin dynamics crucially depends on initial relative phases. 
Another influence on the spin dynamics rates may arise from shot-to-shot varying densities.

In its simplicity the rate-equation model allows to obtain a clear physical picture of the dominant thermodynamical aspects but it lacks coherent spin dynamics which has been reduced to simple rate-equations. 
This procedure seems to be suitable for the discussed regime and may be applied to further problems in this context. Nevertheless 
the detailed treatment of shot-to-shot variations, coherent dynamics, excitations and phase-fluctuations of condensates \cite{Petrov2001a} requires an extended theoretical description.  

Finally we want to point out that the complementary case 
has been studied for $F\!=\!2$ of $^{87}$Rb \cite{Schmaljohann2004a}, where spin dynamics ($\approx$\,10\,ms) is faster than thermalization ($\approx$\,50\,ms) leading first to a steady distribution of condensate spin components which afterwards melt.
  
In conclusion we have reported the experimental realization of a new regime of Bose-Einstein condensation in multi-component spinor condensates at finite temperature. A rate-equation model has been presented to discuss the main thermodynamical aspects. The physics introduced here paves the way towards general new aspects in multi-component quantum gas thermodynamics at finite temperature.

We acknowledge support from the Deutsche Forschungsgemeinschaft in the SPP 1116.

\end{document}